\documentclass[superscriptaddress,showpacs,amssymb,10pt,reprint,aps,prd,longbibliography,nofootinbib,floatfix]{revtex4-1}

\usepackage{graphicx,epsfig,amssymb}
\usepackage{amsmath,amsfonts, times}
\usepackage{bm} 

\usepackage{epstopdf}
\usepackage[linktocpage,colorlinks]{hyperref}
\usepackage[caption=false]{subfig}
\usepackage[usenames]{color}     
\usepackage{natbib}
\usepackage{soul}
\usepackage[utf8x]{inputenc}
\usepackage[usenames]{color}

\definecolor{coolblack}{rgb}{0.0, 0.18, 0.39}
\definecolor{darkred}{rgb}{0.5,0,0}
\definecolor{darkgreen}{rgb}{0,0.5,0}
\definecolor{darkblue}{rgb}{0,0,0.5}
\definecolor{lapislazuli}{rgb}{0.15, 0.38, 0.61}
\definecolor{venetianred}{rgb}{0.78, 0.03, 0.08}
\definecolor{bleudefrance}{rgb}{0.19, 0.55, 0.91}
\definecolor{dogwoodrose}{rgb}{0.84, 0.09, 0.41}
\hypersetup{colorlinks=true, citecolor=darkgreen, linkcolor=darkblue, 
	urlcolor = blue}

\newcommand\numberthis{\addtocounter{equation}{1}\tag{\theequation}}

\def\be{\begin{equation}}
\def\ee{\end{equation}}

\newcommand{\bea}{\begin{eqnarray}}
\newcommand{\eea}{\end{eqnarray}}
\newcommand{\ben}{\begin{enumerate}}
	\newcommand{\een}{\end{enumerate}}
\newcommand{\bi}{\begin{itemize}}
	\newcommand{\ei}{\end{itemize}}

\newcommand{\rt}{r_{\star}}

\def\ga{\mathrel{\raise.3ex\hbox{$>$\kern-.75em\lower1ex\hbox{$\sim$}}}}
\def\la{\mathrel{\raise.3ex\hbox{$<$\kern-.75em\lower1ex\hbox{$\sim$}}}}

\def\l{\left}
\def\r{\right}
\def\be{\begin{equation}}
\def\ee{\end{equation}}

\def\I_M{{I_{\scriptscriptstyle M\times M}}}

\def\be{\begin{equation}}
\def\ee{\end{equation}}
\def\bea{\begin{eqnarray}}
\def\eea{\end{eqnarray}}
\newcommand{\beq}{\begin{eqnarray}}
\newcommand{\eeq}{\end{eqnarray}}
\def\pa{\partial}

\begin{document}
	\title{\large {Schwarzschild-like black holes: Light-like trajectories and massless scalar absorption}}

	\author{Renan B. Magalh\~aes}
	\email{rbmagalhaes22@hotmail.com}
	\affiliation{Programa de P\'os-Gradua\c{c}\~{a}o em F\'{\i}sica, Universidade 
		Federal do Par\'a, 66075-110, Bel\'em, Par\'a, Brazil.}
	
	\author{Luiz C. S. Leite}
	\email{luizcsleite@ufpa.br}
	\affiliation{Programa de P\'os-Gradua\c{c}\~{a}o em F\'{\i}sica, Universidade 
		Federal do Par\'a, 66075-110, Bel\'em, Par\'a, Brazil.}
	\affiliation{Campus Altamira, Instituto Federal do Par\'a, 68377-630, Altamira, Par\'a, Brazil.}
	
	\author{Lu\'is C. B. Crispino}
	\email{crispino@ufpa.br}
	\affiliation{Programa de P\'os-Gradua\c{c}\~{a}o em F\'{\i}sica, Universidade 
		Federal do Par\'a, 66075-110, Bel\'em, Par\'a, Brazil.}

	\begin{abstract}
Black holes are among the most intriguing objects in nature. They are believed to be fully described by General Relativity (GR), and the astrophysical black holes are expected to belong to the Kerr family, obeying the no-hair theorems. Alternative theories of gravity or parameterized deviations of GR allow black hole solutions, which have additional parameters other than mass and angular momentum. We analyze a Schwarzschild-like metric, proposed by Johannsen and Psaltis, characterized by its mass and a deformation parameter. We compute the absorption cross section of massless scalar waves for different values of this deformation parameter and compare it with the corresponding scalar absorption cross section of the Schwarzschild black hole. We also present analytical approximations for the absorption cross section in the high-frequency regime. We check the consistence of our results comparing the numerical and analytical approaches, finding excellent agreement.
	\end{abstract}
	
	\date{\today}
	
	\maketitle

	%%%%%%%%%%%%%%%%%%%%%%%%%%%%%%%%%%%%%%%%%%%%%%
	\section{Introduction}\label{sec:int}
	%%%%%%%%%%%%%%%%%%%%%%%%%%%%%%%%%%%%%%%%%%%%%%
	The no-hair theorems establish a paradigm in black hole (BH) physics: BHs belong to the Kerr family~\cite{costa}, so that the astrophysical BH candidates are fully described by two parameters --- their mass and total angular momentum~\cite{wald}. Although this no-hair paradigm has not been refuted by experimental tests of General Relativity (GR)~\cite{berti2006,berti2016,berti2015,gossan,meidan,isi}, the strong field regime of GR is being put to test~\cite{eht,ligo}, and several works propose deviations to standard GR solutions, by including parameters beyond mass and total angular momentum. 
	
Generalizations of GR solutions have been proposed over the years. Among the approaches are the bumpy BH~\cite{vigeland,collins} and the modified bumpy BH formalism~\cite{yunes}, each of them with advantages and limitations. To surpass the pathologies of bumpy solutions, in $2011$ Johannsen and Psaltis proposed, without the constraint of satisfying the Einstein's equations, a Schwarzschild-like spacetime with an additional parameter~\cite{johannsen}. They applied the Newman-Janis algorithm and obtained a rotating version of their static solution. Since then, many works analyzed the Johannsen-Psaltis (JP) spacetime, e.g. including charge in the BH spacetime~\cite{rahim}, and computing of the photon orbits in this geometry~\cite{pappas}. Generalizations of JP metric~\cite{pani}, and new parametrized solutions~\cite{rezolla,KZ} have also been proposed.	
	
	The decade of 1970 brought a powerful method to analyze the behavior of fields in BH spacetimes, allowing the study of the absorption and scattering of fields by BHs~\cite{fabbri,unruh,sanchez}. The first works investigated the field equations analytically and the range of validity of the solutions were restrict to some limiting (low- or high-frequency) regimes. With numerical methods the absorption and scattering problems became treatable in the whole frequency range. This numerical technique has been extensively revisited along the years~\cite{BODC:2014, MC:2014, CDHO:2015, BC:2016, LDC:2017, LDC:2018, LBC:2019}.

	We study a massless scalar field propagating in the vicinity of a non-spinning JP spacetime. We consider a plane wave impinging from infinity and discuss how the JP deformation parameter influences the absorption by the BH. In the eikonal limit, we use the geodesic equation to obtain high-frequency approximations, comparing them to our numerical results.
	
    The remaining of this paper is organized as follows. In Sec. \ref{sec:JPBH} we present the static Johannsen-Psaltis BH (JPBH) and point out some properties of the corresponding spacetime. In Sec. \ref{sec:class_photon_trajec} we exhibit the photon orbit equation and solve it to determine the classical absorption cross section, which we use to check the consistency of the numerical results in the high-frequency limit. In Sec. \ref{sec:massless_sf} we investigate the massless scalar field, obtaining the radial and angular equations. In Sec. \ref{sec:absorption} we compute numerically the partial and total absorption cross sections of the massless scalar field for the JPBH. We conclude with our final remarks in Sec. \ref{sec:remarks}. We make use of natural units $G = c =\hslash =1$, and signature $(+ - -\, -)$.
    
\section{Non-spinning Johannsen-Psaltis BH} \label{sec:JPBH}
In order to investigate a new class of Schwarzschild-like objects, the JP line element~\cite{johannsen}, namely
\begin{equation}
\label{eq1}
	ds^{2} = \big[1+h(r)\big]\left(f(r)dt^2 - \dfrac{1}{f(r)}dr^2\right) - r^2d\Omega^2,
\end{equation}
has been introduced, where $f(r) \equiv 1 - 2M/r$ and $d\Omega^2 = d\theta^2 + \sin^2\theta\, d\phi^2$ is the line element of an unit sphere. The usual Schwarzschild line element is recovered for a vanishing deformation function $h(r)$. Following Ref.~\cite{johannsen}, we chose $h(r)$ to be a power series of $M/r$, namely
\begin{equation}
		h(r) = \sum_{k = 0}^{\infty}\epsilon_{k}\left(\dfrac{M}{r}\right)^k.
\end{equation}

The asymptotic flatness of the spacetime, together with the experimental data of Lunar Laser Ranging~\cite{boggs} and Cassini experiments~\cite{bertotti} imply in constrains to $h(r)$, reducing it to 
\be
\label{deformationJP}
h_{JP}(r) = \epsilon \dfrac{M^3}{r^3},
\ee
with $\epsilon_{3} \equiv \epsilon$. Hence, the non-spinning JP line element~(\ref{eq1}) takes the form

 \begin{equation}
\label{JP} ds^{2} = \left[1+\epsilon \left(\dfrac{M}{r}\right)^{3}\right]\left[f(r)dt^2 - \dfrac{1}{f(r)}
dr^2\right] - r^2d\Omega^2.
\end{equation}
The spacetime described by the line element \eqref{JP} is asymptotically flat, with spherical symmetry, and with a timelike Killing vector field associated to it. The event horizon location is at $r_{h} = 2M$, as in the case of Schwarzschild spacetime.

We can look for the singularities of the non-spinning metrics associated to the line element \eqref{JP} obtaining the Kre\-tschmann scalar $K =R_{\mu \nu \sigma \rho}R^{\mu \nu \sigma \rho}$, namely
	\begin{align*}
	\label{eq:KS}
	K= &\frac{1}{\left(\frac{M^3 \epsilon }{r^3}+1\right)^6}\Bigg( \frac{48
		M^2}{r^6}-\frac{80 M^4 \epsilon }{r^8} + \frac{432 M^5 \epsilon }{r^9} +
	\\&\frac{184 M^6 \epsilon
		^2}{r^{10}} -\frac{1016 M^7 \epsilon ^2}{r^{11}}+ \frac{1908 M^8 \epsilon
		^2}{r^{12}}+
	\\ &\frac{160 M^9 \epsilon ^3}{r^{13}}-\frac{804 M^{10} \epsilon
		^3}{r^{14}}+ \frac{1632 M^{11} \epsilon ^3}{r^{15}}+\\ &\frac{69 M^{12} \epsilon
		^4}{r^{16}}-\frac{176 M^{13} \epsilon ^4}{r^{17}}+\frac{432 M^{14} \epsilon
		^4}{r^{18}}+\\ &\frac{16 M^{15} \epsilon ^5}{r^{19}}+\frac{16 M^{16} \epsilon
		^5}{r^{20}}+\frac{4 M^{18} \epsilon ^6}{r^{22}}  \Bigg). \numberthis  
\end{align*}

\begin{figure}[h]
\includegraphics[scale=0.5]{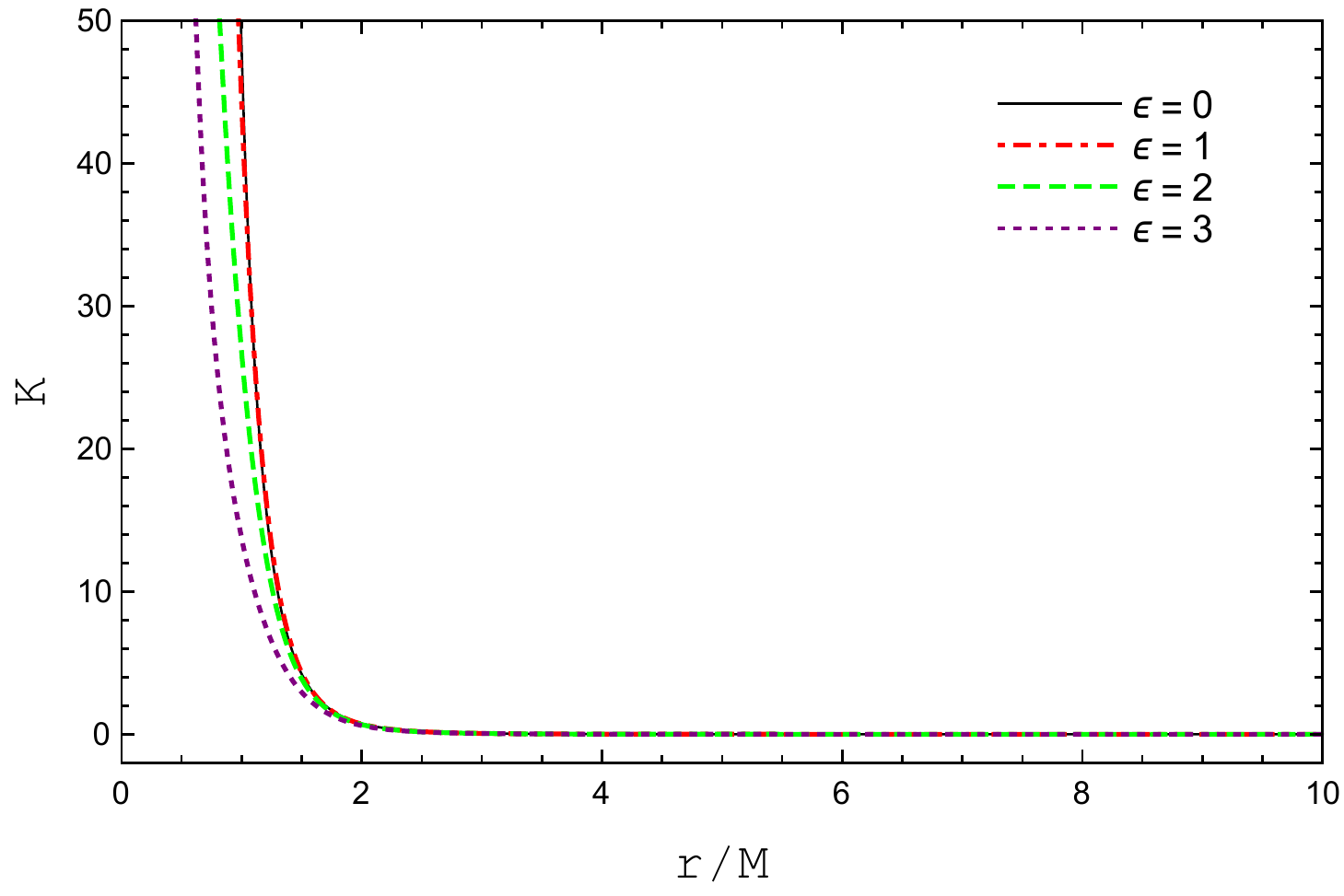}\\
\includegraphics[scale=0.5]{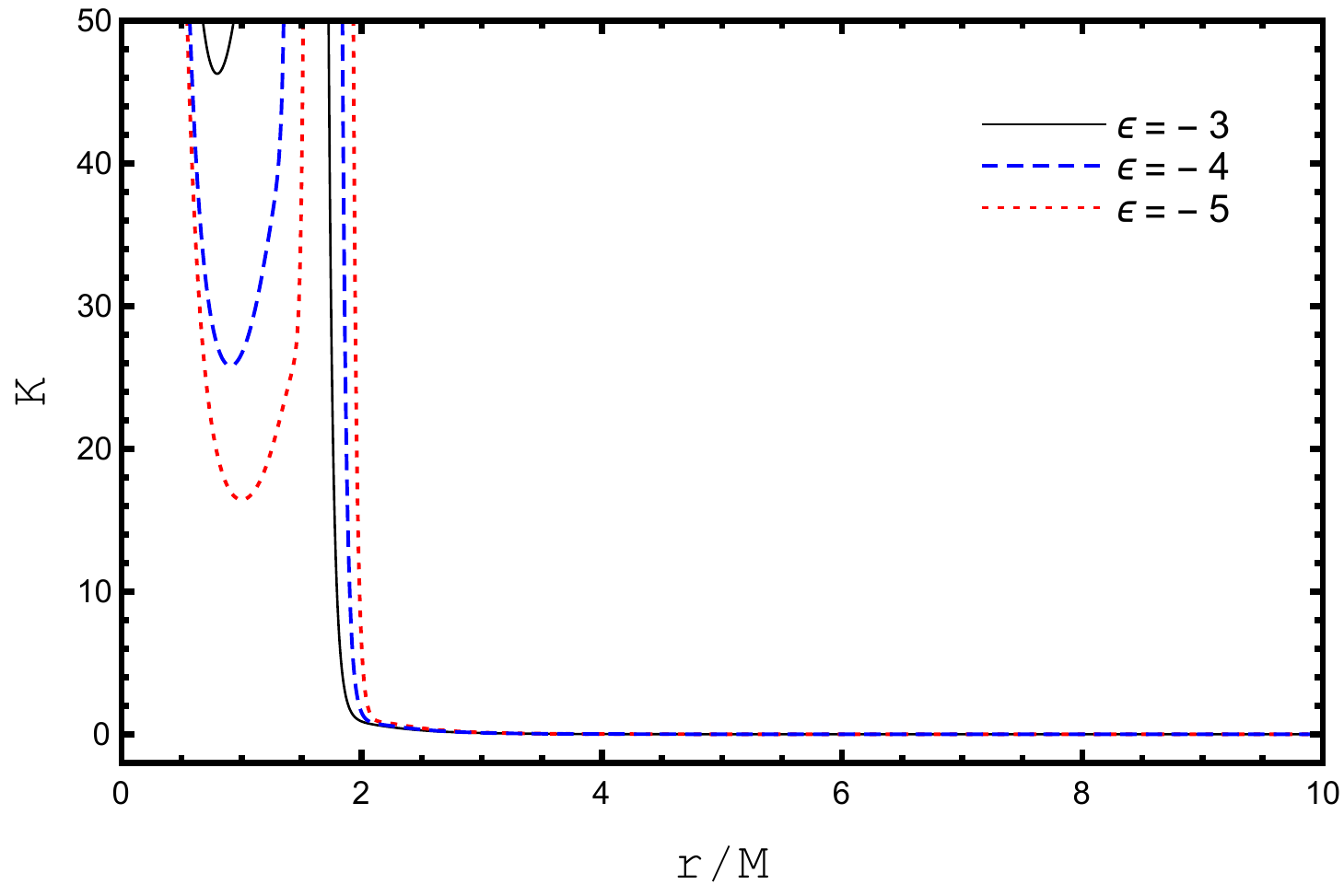}
\caption{Kretschmann scalar, given by Eq.~\eqref{eq:KS}, for some non-negative (top panel) and negative (bottom panel) values of the deformation parameter $\epsilon$.}
\label{PositiveKS}
\end{figure}

This scalar plays an important role in BH physics, enabling us to determine the spacetime singularities. It is important to emphasize that the metric associated to the line element~\eqref{JP} is not required to be a solution of Einstein equations~\cite{johannsen}. By analyzing Eq.~\eqref{eq:KS} for positive values of $\epsilon$, we note that the scalar $K$ diverges at $r=0$, as in Schwarzschild spacetime~($\epsilon=0$). When considering $\epsilon<0$, the Kretschmann scalar diverges for two different radii: the usual $r = 0$ and the additional radius $r = |\epsilon|^{1/3} M$, which defines a surface-like singularity. In Fig.~\ref{PositiveKS}, we plot the Kretschmann scalar for positive (top panel) and negative (bottom panel) values of the deformation parameter $\epsilon$.
%%%%%%%%%%%%%%%%%%%%%%%%%%%%%%%%%%%%%%%%%%%%%%%%%%
%%%%%%%%%%%%%%%%%%%%%%%%%%%%%%%%%%%%%%%%%%%%%%%%%%

\begin{figure}[ht]
\includegraphics[scale=0.5]{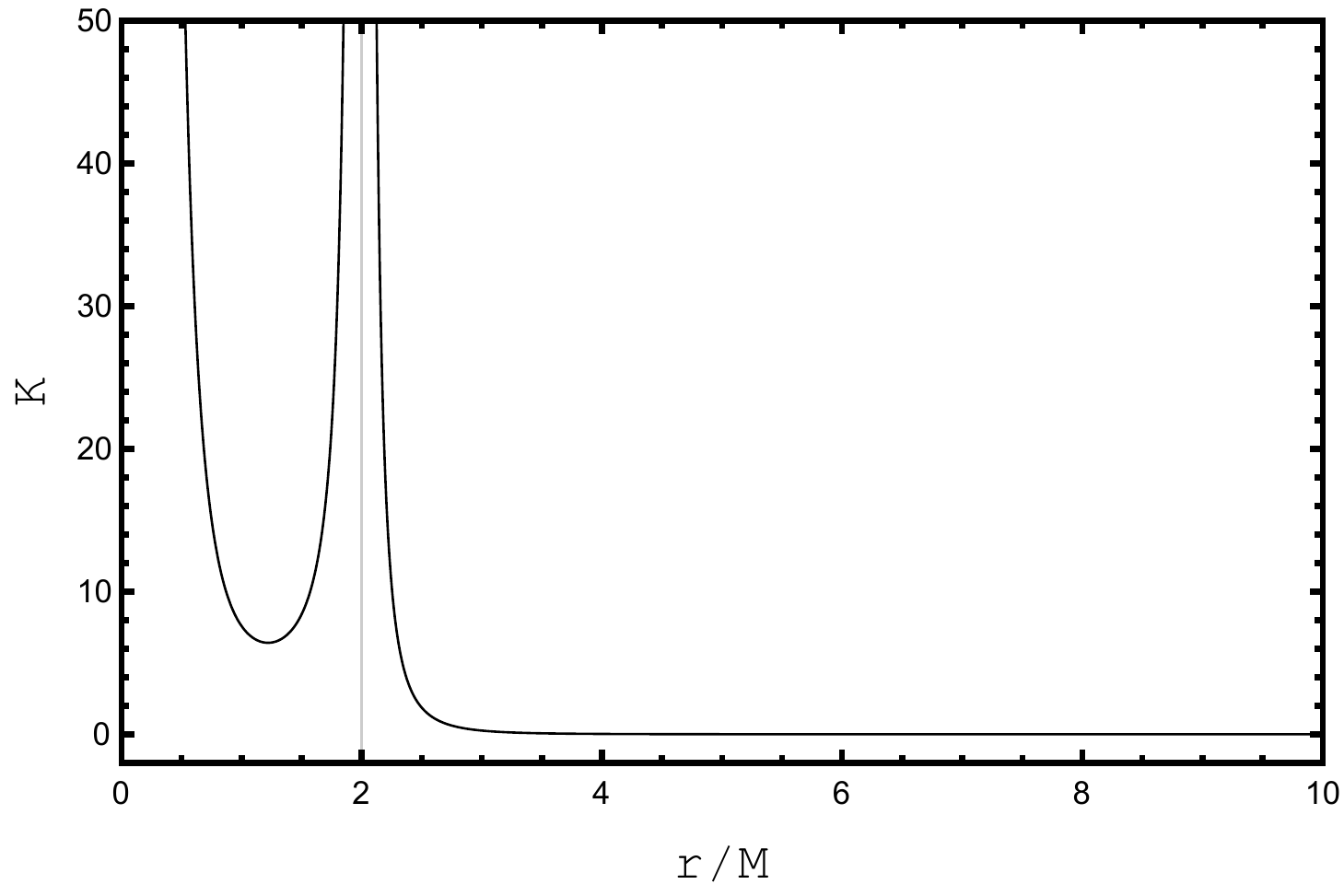}
\caption{Kretschmann scalar for $\epsilon = -8$, with the surface-like singularity located on the event horizon of the JPBH.}
\label{KSnakedsingularity}
\end{figure}

If $\epsilon$ is sufficiently negative, the surface-like singularity lies outside the BH event horizon, being located on the event horizon for $\epsilon = -8$, as it can be seen in Fig~\ref{KSnakedsingularity}, and outside the event horizon for $\epsilon < -8$. Therefore, if the deformation parameter is smaller than $-8$, the line element~\eqref{JP} is associated to a naked singularity, and then violates the cosmic censorship conjecture~\cite{penrose}. Throughout this paper we consider $\epsilon > -8$.

\section{Light-like trajectories} \label{sec:class_photon_trajec}
We can obtain the trajectories of particles in a curved spacetime with covariant metric components $g_{\mu\nu}$, considering the Lagrangian
\be
\label{eq:lagrangianpartilcle}
L = \dfrac{1}{2}g_{\mu\nu}\dot{x}^{\mu}\dot{x}^{\nu},
\ee
where the overdot represents a derivative with respect to the affine parameter. The 4-velocity $\dot{x}^{\mu}$ is normalized to unity for massive particles ($\dot{x}^{\mu}\dot{x}_{\mu}=1$) and has vanishing norm ($\dot{x}^{\mu}\dot{x}_{\mu}=0$) for massless particles, so that for photons we have
\be
\label{eq:lagrangianphoton}
L=0.
\ee

\begin{figure}[h]
\includegraphics[scale=0.5]{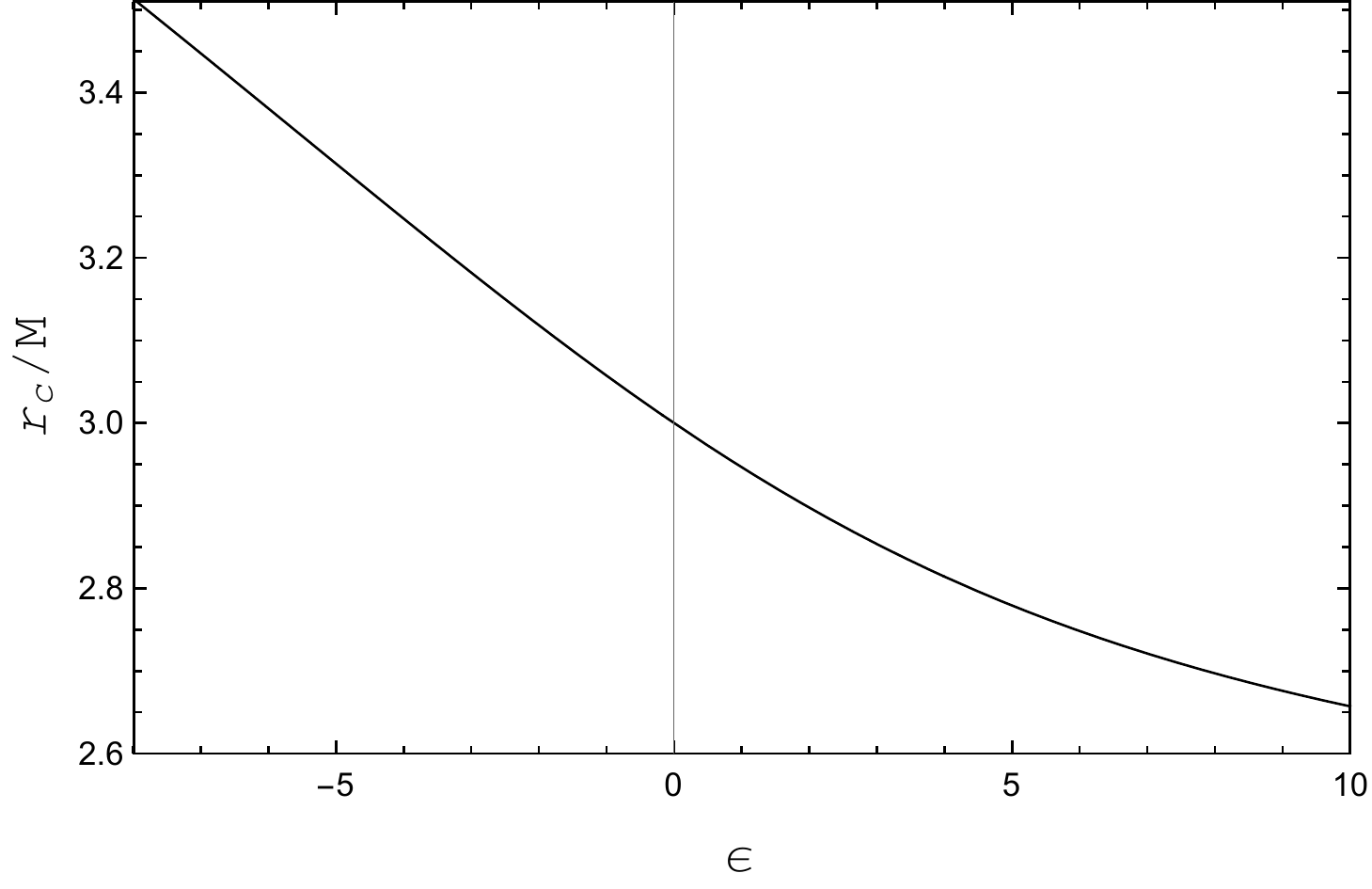}
\caption{Light sphere location ($r_c$), plotted as a function of the JPBH deformation parameter $\epsilon$.}
\label{lightring}
\end{figure}

Since the line element~\eqref{JP} is spherically symmetric, without loss of generality we may evaluate Eq.~\eqref{eq:lagrangianphoton} in the equatorial plane $(\theta = \pi/2)$, and obtain the orbit equation for massless particles, given by
    \begin{equation}
   \label{eq:orbit} \left(\dfrac{d u}{d \phi}\right)^2 = -u^2\left(\dfrac{1-2 M u}{1+ \epsilon M^3 u^3}\right)+\dfrac{1}{b^2\left(1+\epsilon M^3u^3\right)^2},
\end{equation}
where $u(\phi) \equiv 1/r(\phi)$ and $b$ is the impact parameter~\cite{wald}. Equation~\eqref{eq:orbit} allows an unstable circular orbit (also called light ring or light sphere) at $r=r_{c}$. To determine the critical radius $r_{c}$ for different values of the deformation parameter $\epsilon$, we impose that $du/d\phi = 0$, and $d^2u/d\phi^2 = 0$. We plot the light sphere radius ($r_{c}$) for several JPBHs in Fig.~\ref{lightring}.

\begin{figure}[h]
 	\includegraphics[scale= 0.5]{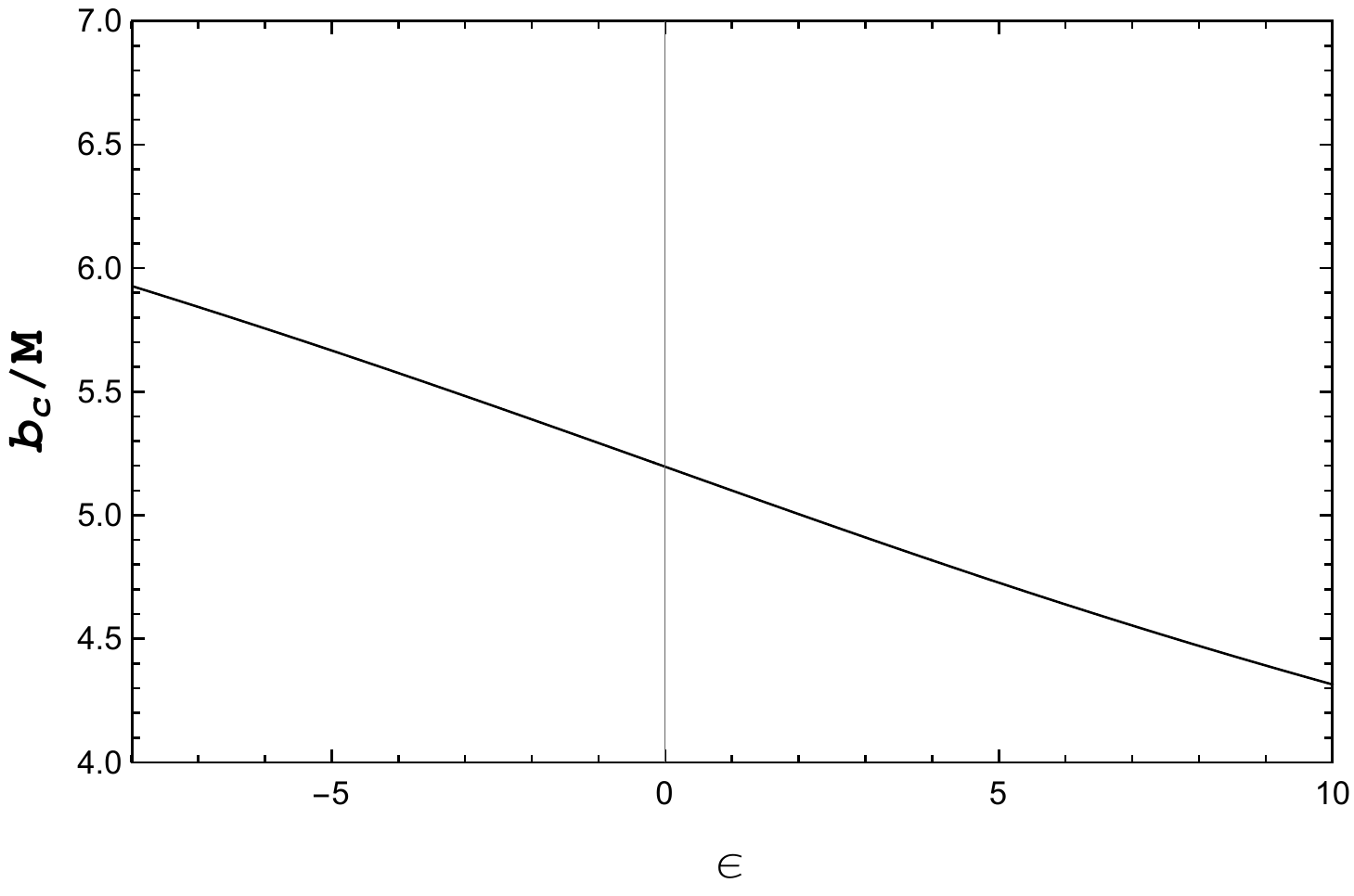}
    \caption{Critical impact parameter $b_{c}$, plotted as a function of $\epsilon$ for JPBHs.}
\label{bc-sigma}
\end{figure}

There is a critical impact parameter $b = b_{c}$, so that any photon, with impact parameter equal to $b_{c}$, incoming from spatial infinity stays trapped in the light sphere at $r=r_{c}$. The expression for the critical impact parameter $b_{c}$ as a function of the JP parameter $\epsilon$ is cumbersome and we prefer not to show it here. We plot the critical impact parameter for different JPBHs in Fig.~\ref{bc-sigma}. 

\begin{figure}[h]
\includegraphics[scale= 0.4]{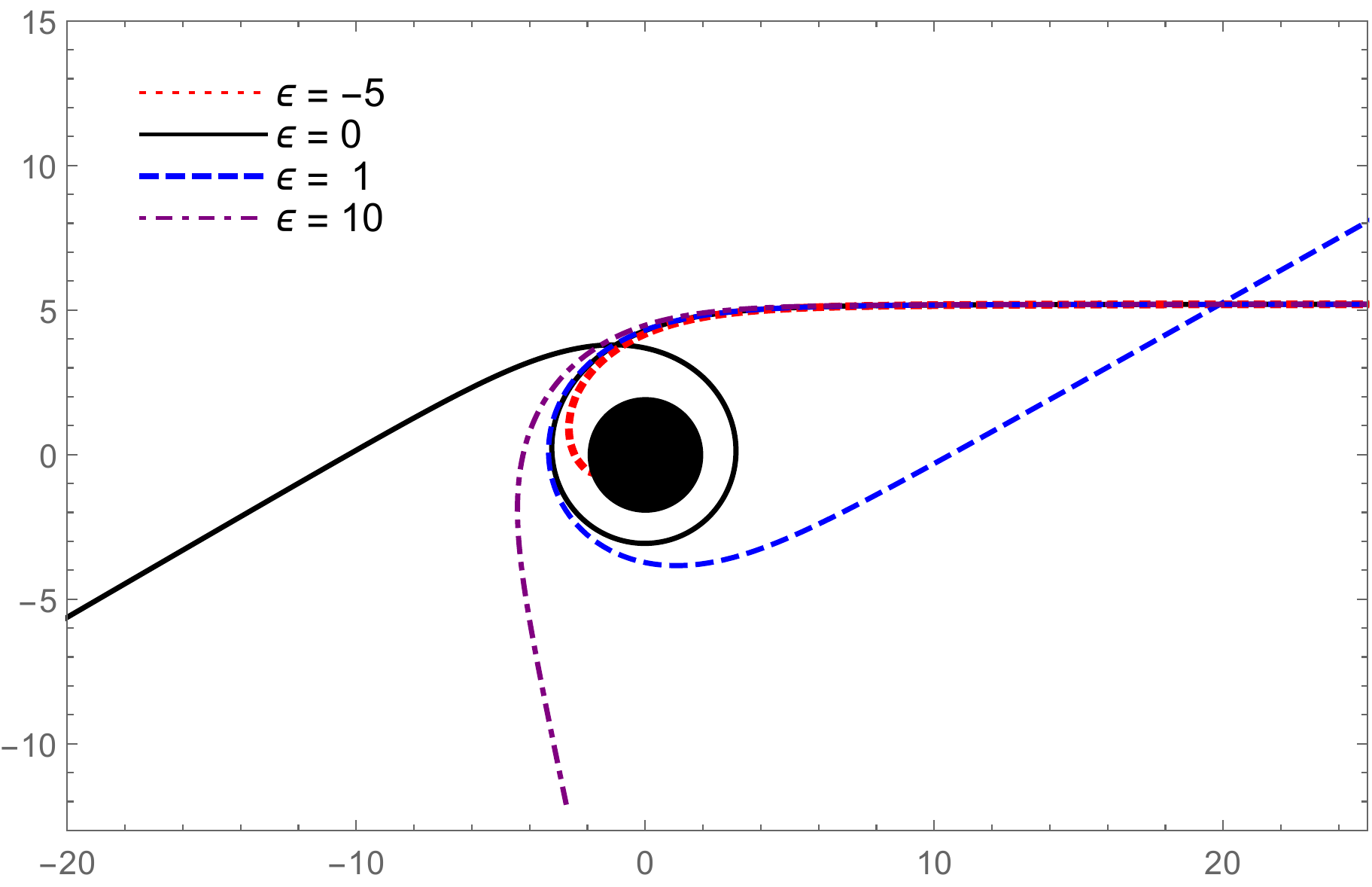}
\caption{Null geodesics with $b = 5.2M$, considering different choices of $\epsilon$. We note that the JPBH has a stronger influence in photon trajectories for smaller values of the deformation parameter.}
\label{scatteps}
\end{figure}

In Fig.~\ref{scatteps} we plot the photon trajectories for different choices of the JP parameter $\epsilon$, fixing the impact parameter $b = 5.2M$. We note that, light rays are less deflected by JPBHs with a bigger deformation parameter.

\begin{figure}[h]
\centering
		\includegraphics[scale= 0.5]{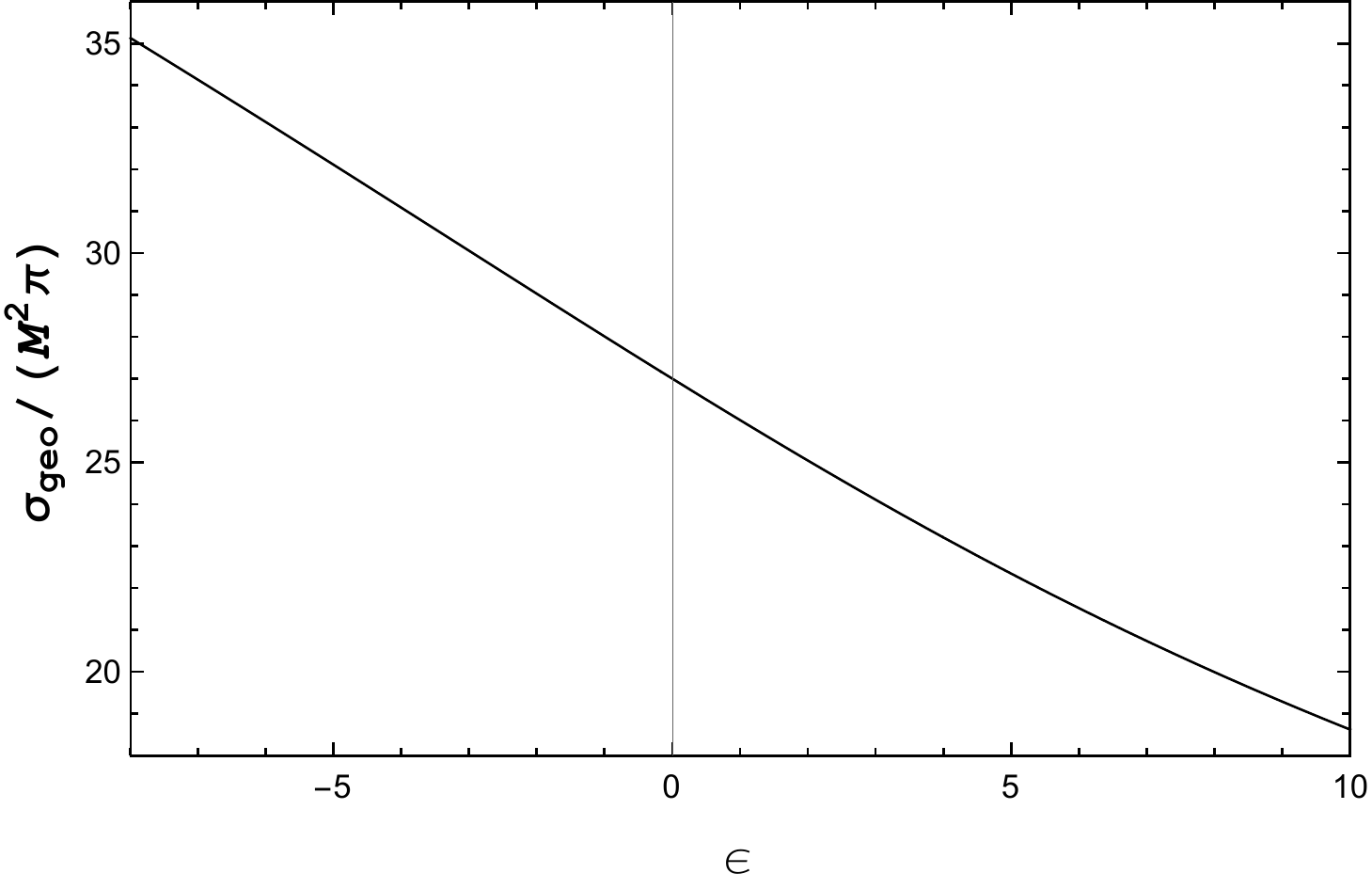}\\
			\includegraphics[scale=0.5]{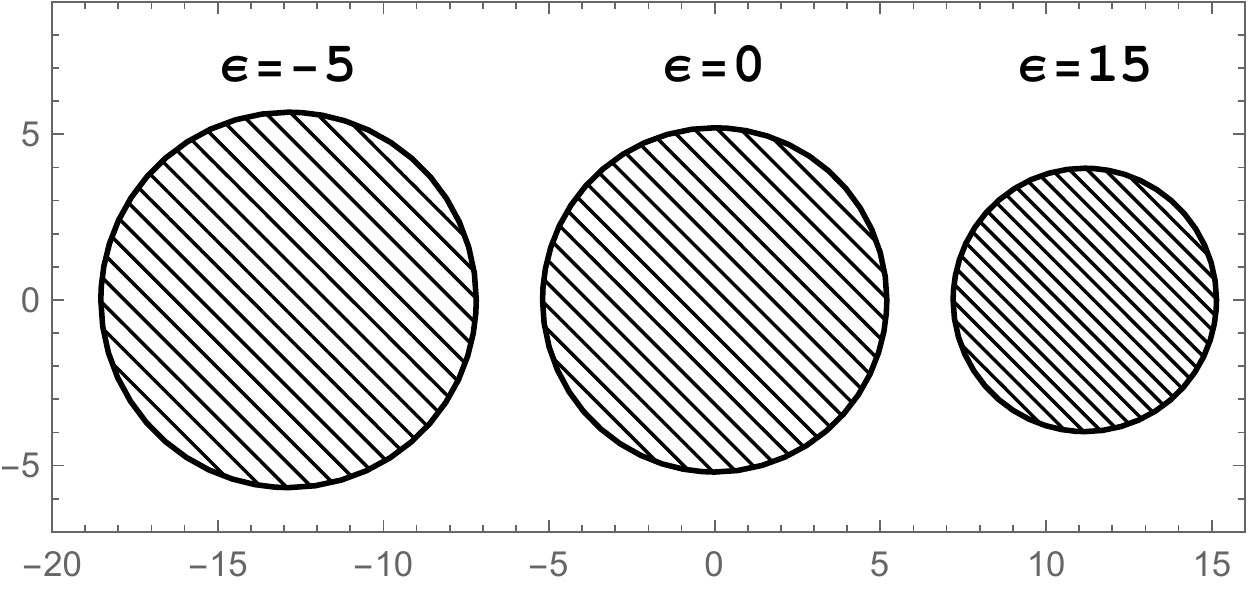}
		\caption{Top: Classical absorption cross section, as a function of $\epsilon$. Bottom: Geometrical representation of the classical absorption cross section for different choices of $\epsilon$.}
		\label{ca-sigma}
	\end{figure}

The geometrical (or classical) absorption cross section is given by the area of a disk with radius $b_{c}$, namely
\be
\label{eq:clabs}
\sigma_{geo} = \pi b^2_{c}.
\ee
From Fig.~\ref{ca-sigma}  
we notice that as the deformation parameter of the JPBH increases, 
the geometrical absorption cross section diminishes.

	In Sec.~\ref{sec:absorption} we use this classical analysis as a consistency check of our numerical results in the high-frequency regime.
%%%%%%%%%%%%%%%%%%%%%%%%%%%%%%%%%%%%%%%%%%%%%%%%%%
%%%%%%%%%%%%%%%%%%%%%%%%%%%%%%%%%%%%%%%%%%%%%%%%%%
\section{Massless scalar field dynamics}\label{sec:massless_sf}
In order to study the absorption of massless scalar waves by Schwazschild-like BHs, we consider a spin-0 field governed by the (massless) Klein-Gordon equation,
	\be
	(-g)^{-1/2}\pa_\mu(g^{\mu\nu}\sqrt{-g}\pa_\nu\Psi)=0,\label{eq:kge}
	\ee
	with $\pa_\mu\equiv\pa/\pa x^\mu$ and $g^{\mu\nu}$ being the contravariant metric components. For the JPBH, the metric determinant $g$ is
	\be
\label{eq:metricdet}
g = -\dfrac{\left(r^3 + M^3\epsilon\right)^2}{r^2}\sin^{2}\theta.	
	\ee
	
	Due to the spherical symmetry, the massless scalar field~$\Psi$ can be conveniently decomposed as follows
	\be
	\Psi(x^\mu)=\frac{\psi_{\omega l}(r)}{r}Y_{lm}(\theta,\,\phi)e^{-i\omega t},
	\label{eq:decom}
	\ee
	where $Y_{lm}(\theta,\phi)$ are the spherical harmonics with eigenvalues $l(l+1)$. By substituting Eq.~\eqref{eq:decom} in Eq.~\eqref{eq:kge} we obtain the ordinary differential equation for the radial function $\psi_{\omega l}(r)$, namely
	\be
	f(r)\frac{d}{dr}\l[f(r)\frac{d\psi_{\omega l}}{dr}\r]+\l[\omega^2-V_l(r)\r]\psi_{\omega l}=0,\label{eq:radialeq}
	\ee
	where $V_l$ is the effective potential, given by
	\be
	\label{eq:effpotential}
	V_l(r)\equiv \frac{f(r)}{r}\frac{df}{dr}+f(r)\frac{l(l+1)}{r^2}\left[1+\epsilon\left(\dfrac{M}{r}\right)^3\right].
	\ee
	
	Compared with the well-known Schwarzschild case, the effective potential presents an extra $\epsilon$ dependent term. We note that the only deformed contribution in Eq.~\eqref{eq:effpotential} is multiplied by the eigenvalue $l(l+1)$ of the spherical harmonics, so that the lower mode contribution $(l = 0)$ coincides with the Schwarzschild one for any JPBH. 
	
	By introducing the tortoise coordinate $\rt$, namely
	\be
	\label{eq:tortoise}
	\rt\equiv\int \frac{dr}{f(r)},
	\ee
	we can rewrite Eq.~\eqref{eq:radialeq} as a Schrödinger-like equation:
	\be
	\frac{d^2\psi_{\omega l}}{d\rt^2}+\l[\omega^2-V_l\r]\psi_{\omega l}(\rt)=0.\label{eq:radialeqtor}
	\ee

\begin{figure}[ht]
	\includegraphics[scale=0.55]{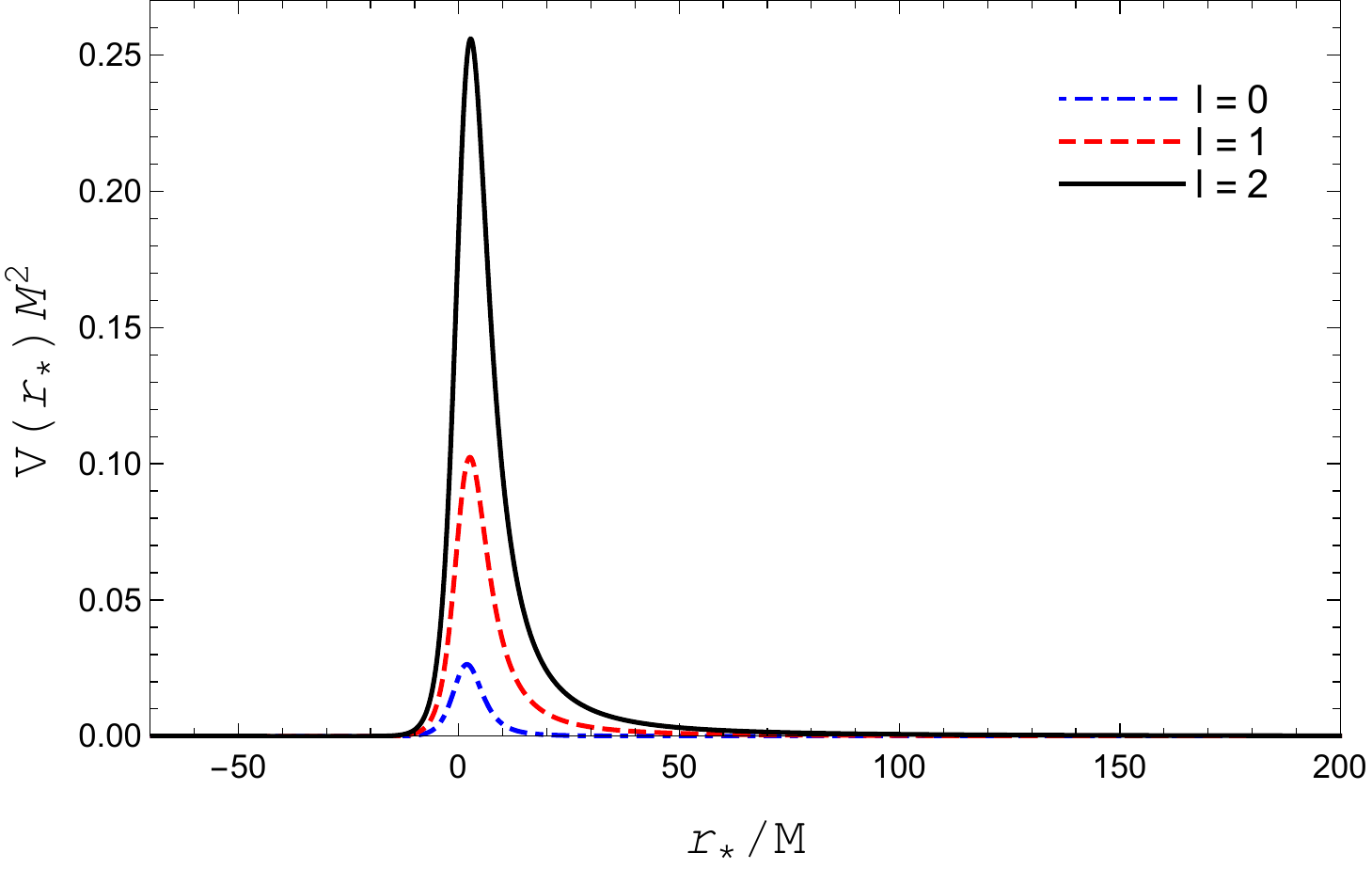}
\caption{Effective potential with $\epsilon = 1$, for modes $l=0$ (blue dot-dashed line), $l=1$ (red dashed line) and $l=2$ (black solid line).}	
	\label{fig:RWpotentiall012}
	\end{figure}

\begin{figure}[t]
	\includegraphics[scale=0.55]{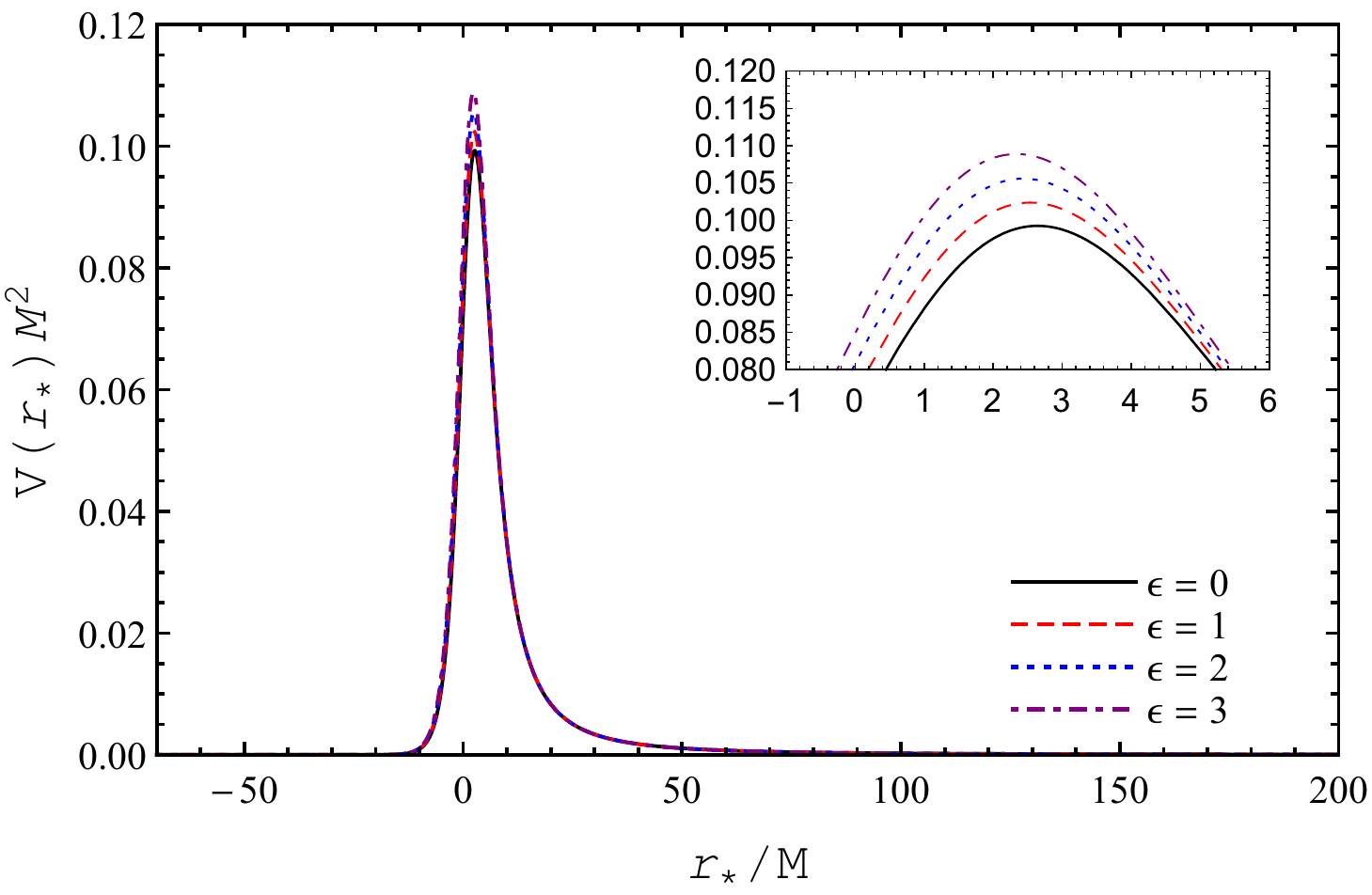}
	\includegraphics[scale=0.55]{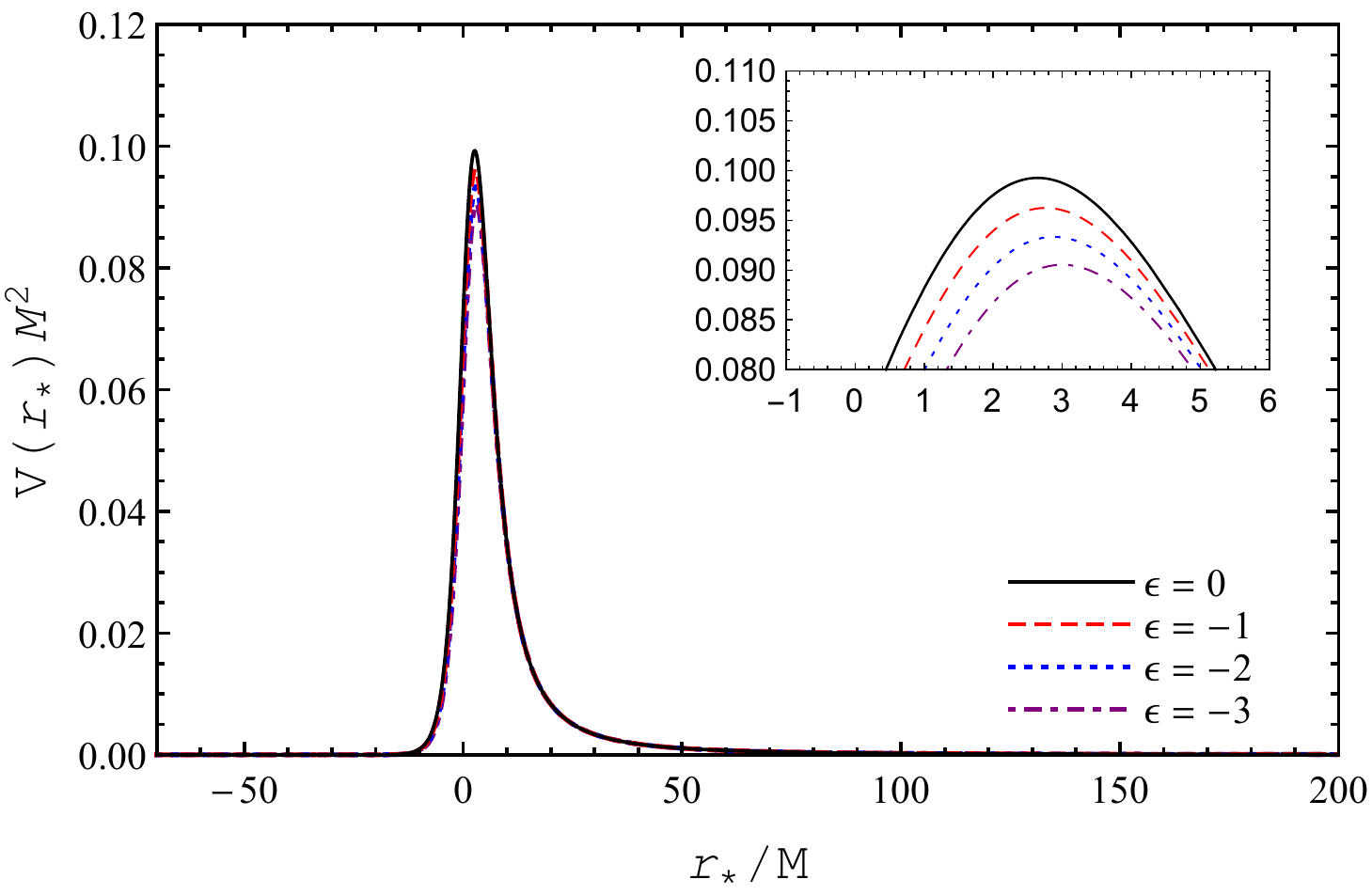}
\caption{Effective potential, given by Eq.~\eqref{eq:effpotential}, with $l=1$, for non-negative (top panel) and non-positive (bottom panel) values of $\epsilon$.}	
	\label{fig:RWpotentialPel1}	
	\end{figure}

We notice in Fig.~\ref{fig:RWpotentiall012} that for a fixed deformation parameter $\epsilon$, the peak of the effective potential is bigger as we increase the mode number $l$. For a fixed mode number $l$, the peak of the effective potential increases as we increase the deformation parameter $\epsilon$, as it can be seen in Fig.~\ref{fig:RWpotentialPel1}.
	
	In order to investigate the absorption of massless scalar waves, we seek for solutions of Eq.~\eqref{eq:radialeqtor} that are purely incoming waves at the event horizon and a composition of ingoing and outgoing waves at spatial infinity, i.e, that satisfy the following boundary conditions:
	\be
	\psi_{\omega l}(\rt)\sim\l\{
	\begin{array}{ll}
		e^{-i \omega \rt}+{\cal R}_{\omega l }e^{i \omega \rt},&\rt\to+\infty, \\
		{\cal T}_{\omega l }e^{-i\omega \rt},& \rt\to-\infty.
	\end{array}\r.\label{eq:inmodes}
	\ee 
	The reflection and transmission coefficients are related to ${\cal R}_{\omega l }$ and ${\cal T}_{\omega l }$, respectively. Due to the flux conservation, the following relation is satisfied:
	\be
	|{\cal R}_{\omega l }|^2+|{\cal T}_{\omega l }|^2=1.
	\ee
	%%%%%%%%%%%%%%%%%%%%%%%%%%%%%%%%%%%%%%%%%%%%%%%%%%
	%%%%%%%%%%%%%%%%%%%%%%%%%%%%%%%%%%%%%%%%%%%%%%%%%%
	%%%%%%%%%%%%%%%%%%%%%%%%%%%%%%%%%%%%%%%%%%%%%%%%%
	%%%%%%%%%%%%%%%%%%%%%%%%%%%%%%%%%%%%%%%%%%%%%%%%%
	%%%%%%%%%%%%%%%%%%%%%%%%%%%%%%%%%%%%%%%%%%%%%%%%%
	\section{Absorption}\label{sec:absorption}
	The absorption cross section can be defined as the ratio between the number of particles absorbed by the black hole and the incident particle flux. One can use the partial waves method to obtain the total absorption cross section of a scalar field \cite{CDO}, namely:
  \begin{equation}
  \label{abs}
     \sigma = \sum^{\infty}_{l=0}\sigma_{l},
 \end{equation}
where $\sigma_{l}$ are the partial absorption cross sections, defined as
\be
\label{partialabs}
\sigma_{l} \equiv \dfrac{\pi}{\omega^2}(2l+1)\Gamma_{l}(\omega).
\ee
The $\Gamma_{l}$ in Eq.~\eqref{partialabs} are the greybody factors, which are related to the absorption probability~\cite{ahmed}. Considering the asymptotic expansion~\eqref{eq:inmodes}, the greybody factors may be written as
\begin{equation}
\label{greybody}
\Gamma_{l} = |{\cal T}_{\omega l}|^2.
\end{equation} 
The total absorption cross section is well-known for static BHs in the low- and high-frequency regimes. In the low-frequency regime Higuchi found that the massless scalar absorption cross section goes to the area of the event horizon for any stationary BH spacetime~\cite{higuchi}. In the high-frequency regime the absorption cross section oscillates around the geometrical capture cross section~\eqref{eq:clabs}, and we can use the analytical sinc approximation (cf. Sec.~\ref{subsec:sinc}) to determine the high-frequency behaviour. 

In the remaining of this section we compute the total absorption cross section using both the sinc approximation and the numerical method discussed in Ref.~\cite{dolan}.

\subsection{Sinc approximation}\label{subsec:sinc}
 Sanchez proposed, to the Schwarzschild BH case, the following analytical approximation for the total scalar absorption cross section, in the high-frequency regime:
\begin{equation}
\label{sanchpaper}
\sigma_{\text{San}} = \dfrac{27\pi}{4} - \dfrac{A}{\omega r_{h}}\sin \pi(3\sqrt{3})(\omega r_{h} + B), 
\end{equation} 
where $A = 1.41\sim\sqrt{2}$ and $B<10^{-4}$ give the best fit. This approximation has been obtained analytically in Ref.~\cite{folacci} for Schwarzschild BHs, in the high-frequency regime, based in an analytical extension of the greybody factor, and summing over $l$-modes using the Poisson sum formula. Following Ref.~\cite{folacci}, one can write the approximation for the scalar absorption by a Schwarzschild BH as:
 \be
 \sigma_{abs} = \sigma_{geo} + \sigma_{RP} + \mathcal{O}\left(\dfrac{1}{\omega^2}\right), 
 \label{eq:sanchez}
 \ee
 where $\sigma_{geo}$ is given by Eq.~\eqref{eq:clabs}, and $\sigma_{RP}$ is a sum over Regge poles, given by
 \be 
 \sigma_{RP} = - \dfrac{4\pi^2}{\omega^2}\text{Re}\left(\sum^{+\infty}_{n=1} \dfrac{\lambda_{n}(\omega)\gamma_{n}(\omega)\,e^{i\pi(\lambda_{n}(\omega) - 1/2)}}{\sin[\pi(\lambda_{n}(\omega) - 1/2)]}\right).
\label{eq:absosc} 
 \ee
 Here $\lambda_{n}$ are the Regge poles and $\gamma_{n}$ are the residues of the greybody factor.

\begin{figure}[h]
\includegraphics[scale=0.5]{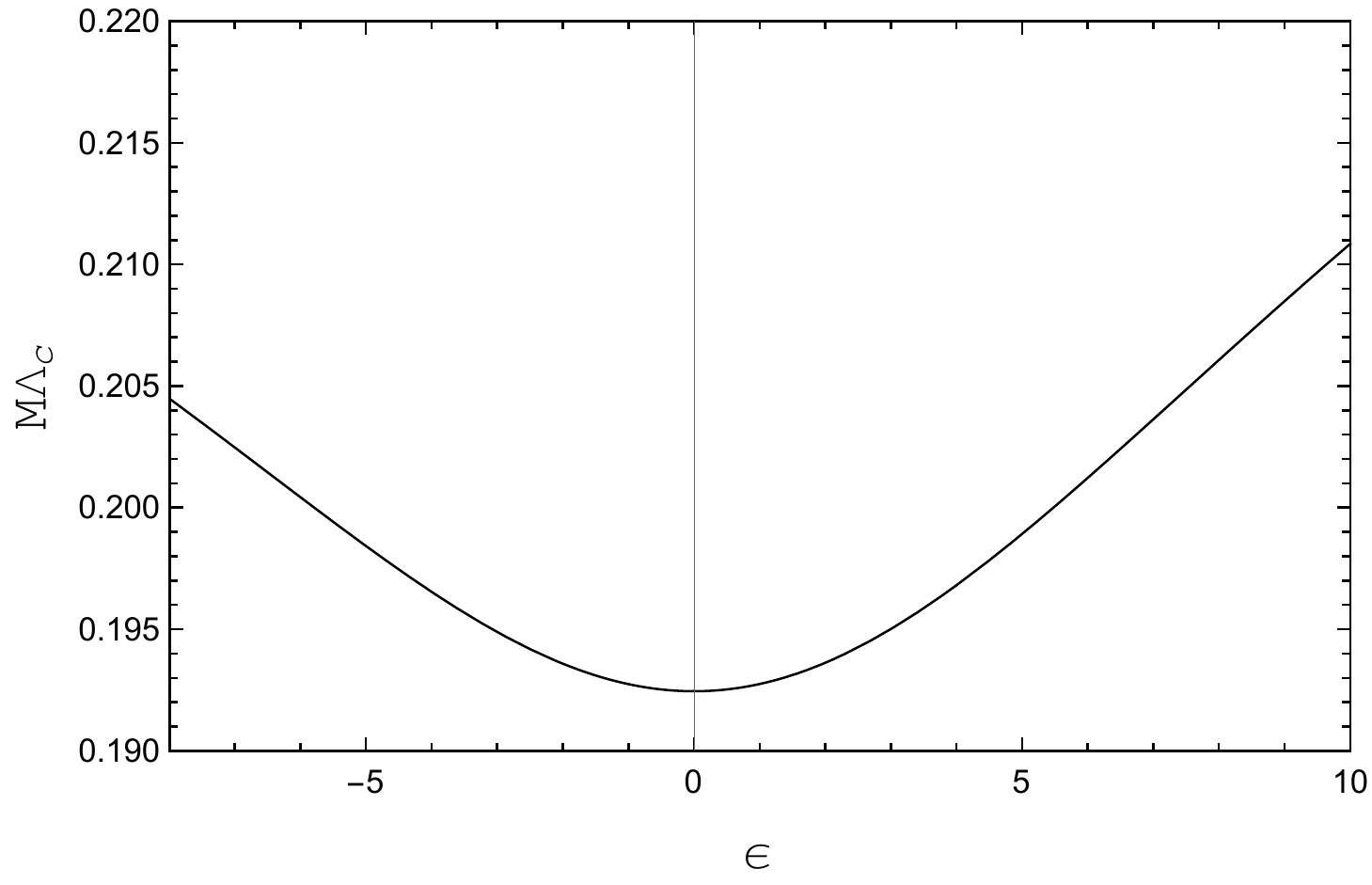}\\
\includegraphics[scale=0.5]{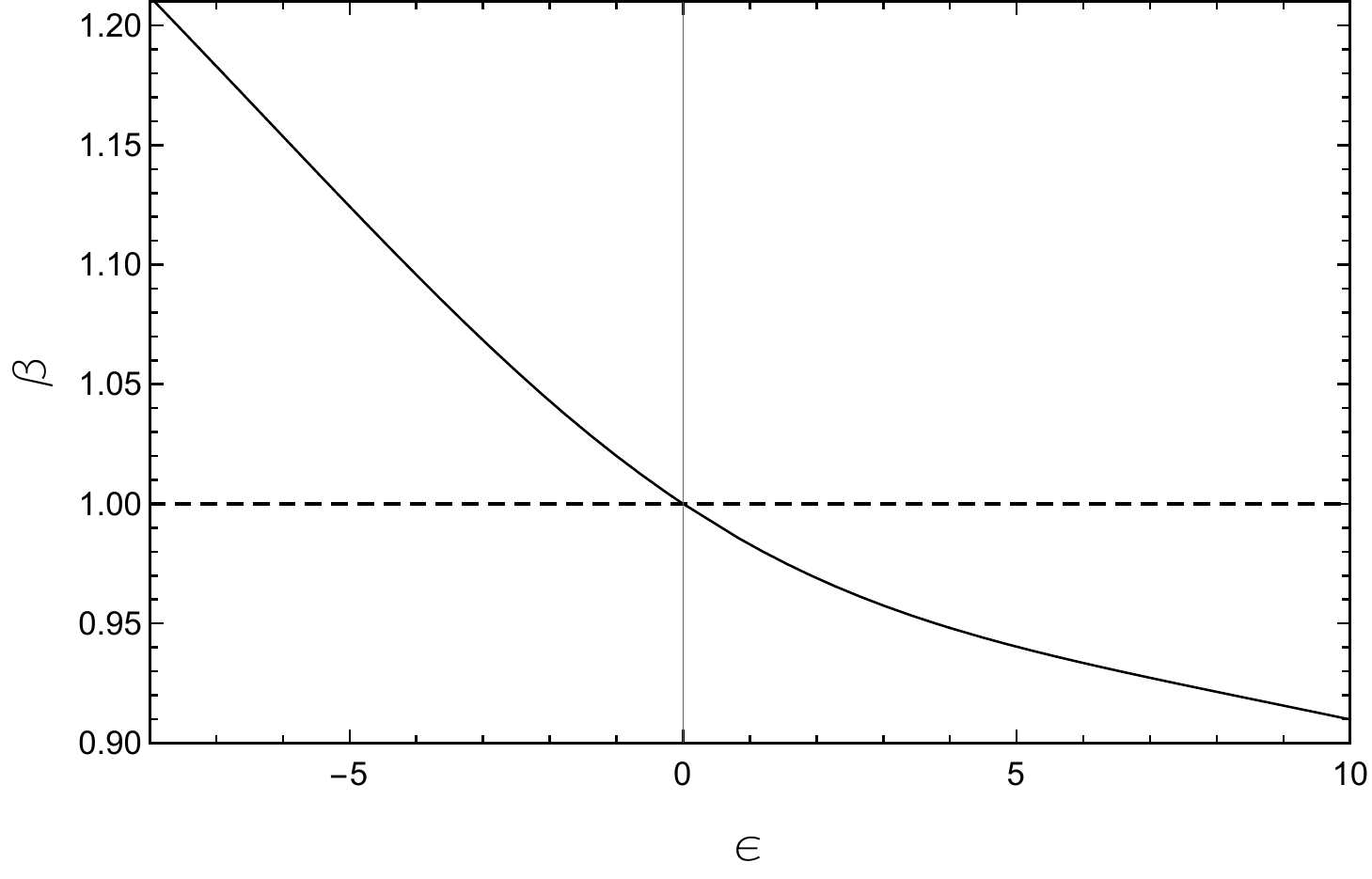}
\caption{Top: Lyapunov exponent $\Lambda_{c}$ at the unstable circular orbit radius $r_{c}$, as a function of $\epsilon$. Bottom: $\beta$ factor as a function of $\epsilon$. When the deformation parameter vanishes the $\beta$ factor is equal to 1.}
\label{lyapunov}
\end{figure}

D\'ecanini et. al. generalized the Sanchez approximation \eqref{sanchpaper}, for an arbitrary static and spherically symmetric BH, obtaining that the total scalar absorption cross section in the high frequency limit, in a four dimensional spacetime, is~\cite{folacci}
\be
 \sigma^{hf}_{abs} = \sigma_{geo}\left[1 - 8\pi\beta\,e^{-\pi\beta}\text{sinc}\left(\dfrac{2\pi\omega}{\Omega_{0}}\right)\right],
\label{eq:folacci}
\ee
where the sine cardinal is defined as $\text{sinc}(z) \equiv \sin(z)/z$. The $\Omega_{0}$ is the orbital frequency, which is the inverse of the critical impact parameter $b_{c}$~\cite{cardoso} plotted in Fig.~\ref{bc-sigma}. The $\beta$ factor is related to the Lyapunov exponent $\Lambda$, by~\cite{raffaelli}
\begin{equation}
\beta =\dfrac{\Lambda_{c}}{\Omega_{0}},
\end{equation}
where $\Lambda_{c}$ is the Lyapunov exponent at the unstable circular orbits radius $r_{c}$, as introduced in Ref.~\cite{cardoso}. The Lyapunov exponent $\Lambda_{c}$ and the $\beta$ factor are plotted in Fig.~\ref{lyapunov}. Their expressions are cumbersome for JPBH, and we prefer not to show them here. If the deformation parameter vanishes, $\beta=1$ (cf. bottom panel of Fig.~\ref{lyapunov}), and we recover the for Schwarzschild case. 

\subsection{Numerical method}
In order to obtain the total absorption cross section in the whole frequency range, we solve the radial equation~\eqref{eq:radialeqtor} numerically, imposing the boundary conditions~\eqref{eq:inmodes}. The integration is made from close enough to the event horizon to a sufficiently large value of the radial coordinate. We compare the radial solution obtained numerically with the asymptotic solution~\eqref{eq:inmodes}, finding the reflection and transmission coefficients. For more details of the numerical method and the convergence of the solution, see Ref.~\cite{dolan}.

\begin{figure}[h]
\includegraphics[scale=0.5]{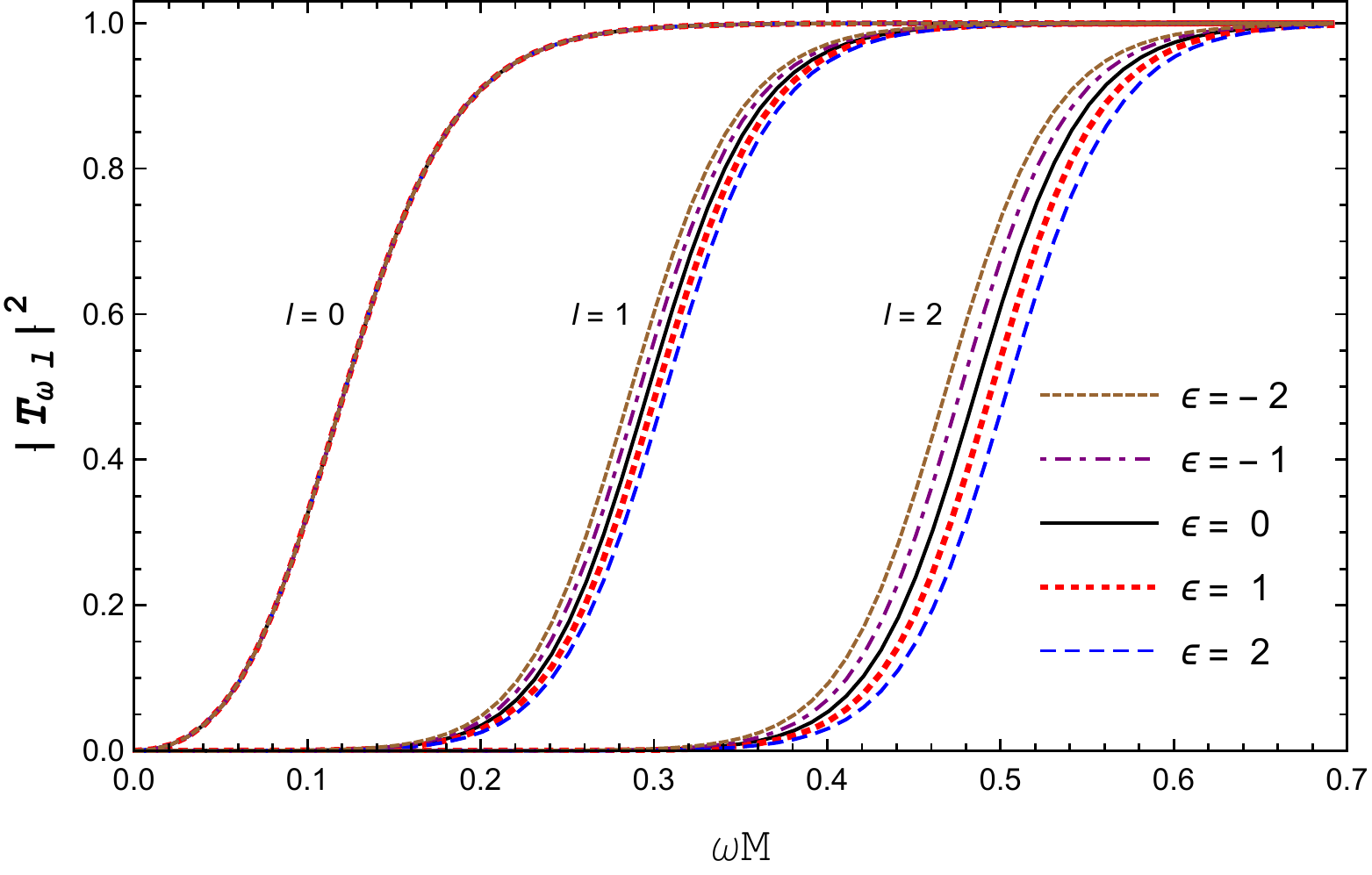}
\caption{Transmission coefficients for the first three modes ($l=0$, $l=1$ and $l=2$), for different choices of the deformation parameter ($\epsilon = -2$, $-1$, $0$, $1$ and $2$). The transmission coefficients for $l=0$ are the same for any value of the deformation parameter, while for non-vanishing angular momentum they are different, and the difference increases for higher values of $l$.}
\label{transmission}
\end{figure}

\begin{figure}[h]
\includegraphics[scale= 0.53]{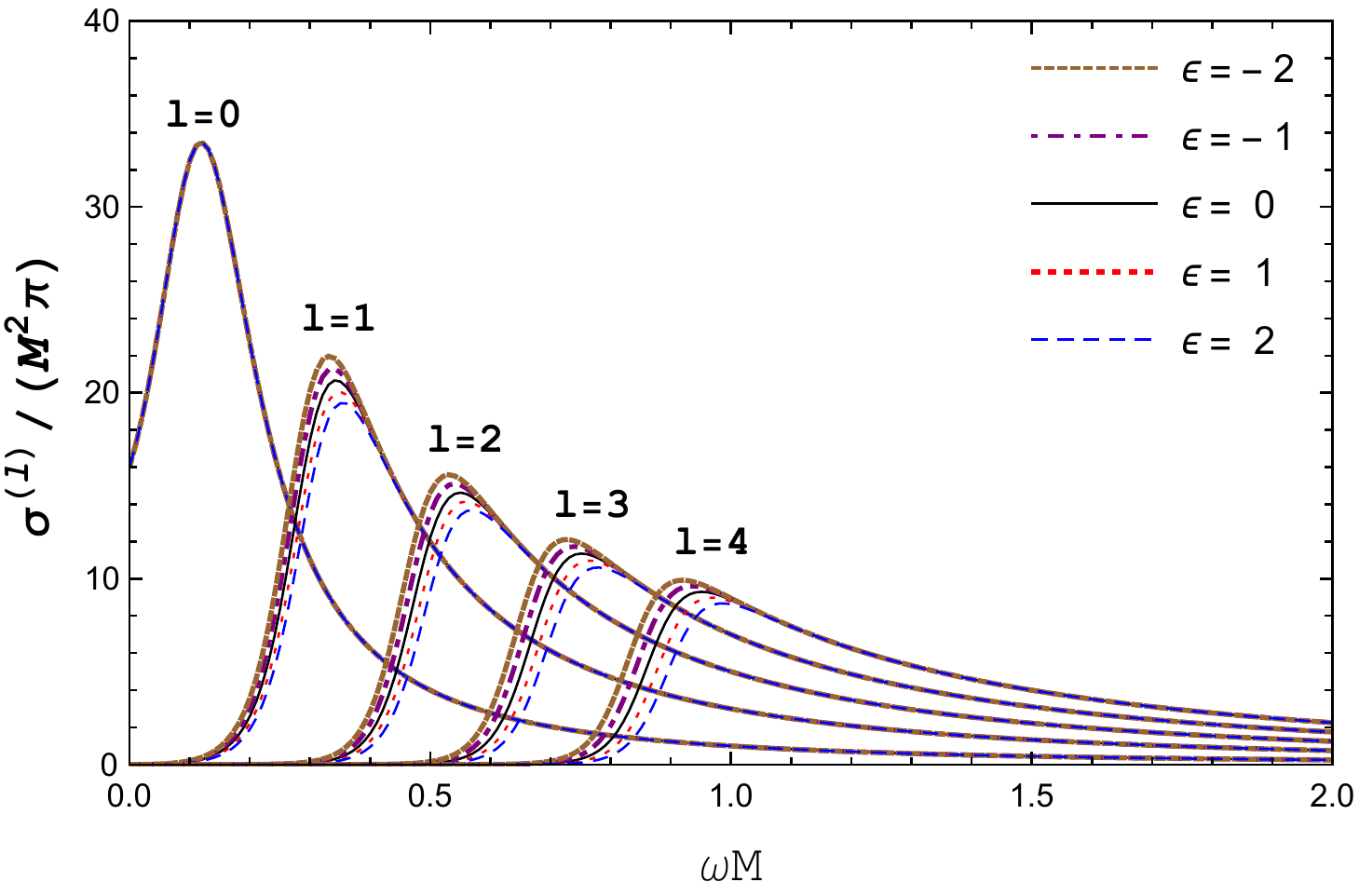}
\caption{Partial absorption cross sections for different choices of the deformation parameter $\epsilon$.}
\label{partial}
\end{figure}

\begin{figure}[h]
\includegraphics[scale= 0.5]{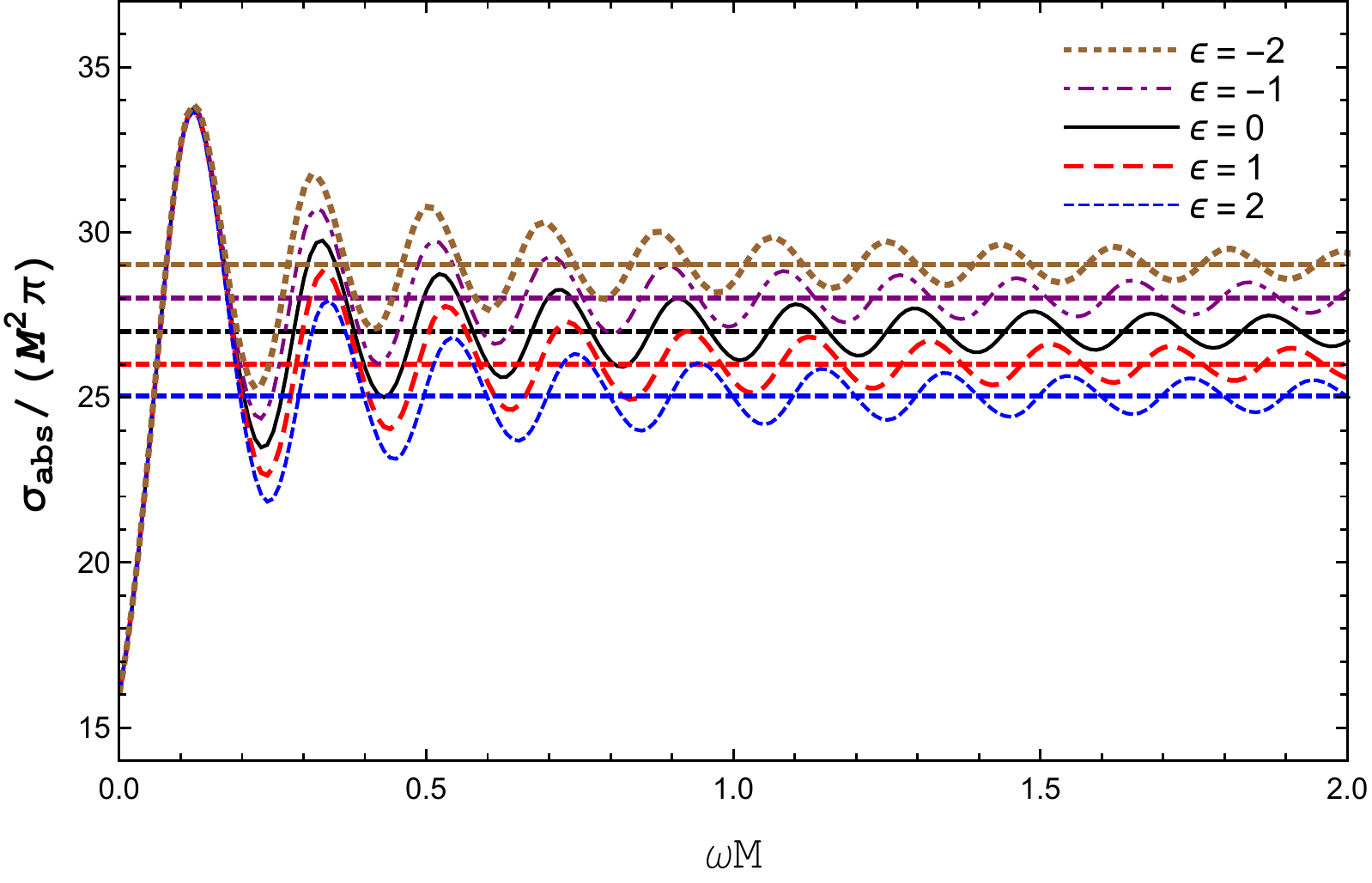}
\caption{Total absorption cross section of Schwarzschild-like objects for different values of $\epsilon$. The horizontal dashed lines represent the geometrical absorption cross sections in each case.}
\label{Schwarzschild}
\end{figure}

\begin{figure}[ht]
\includegraphics[scale= 0.5]{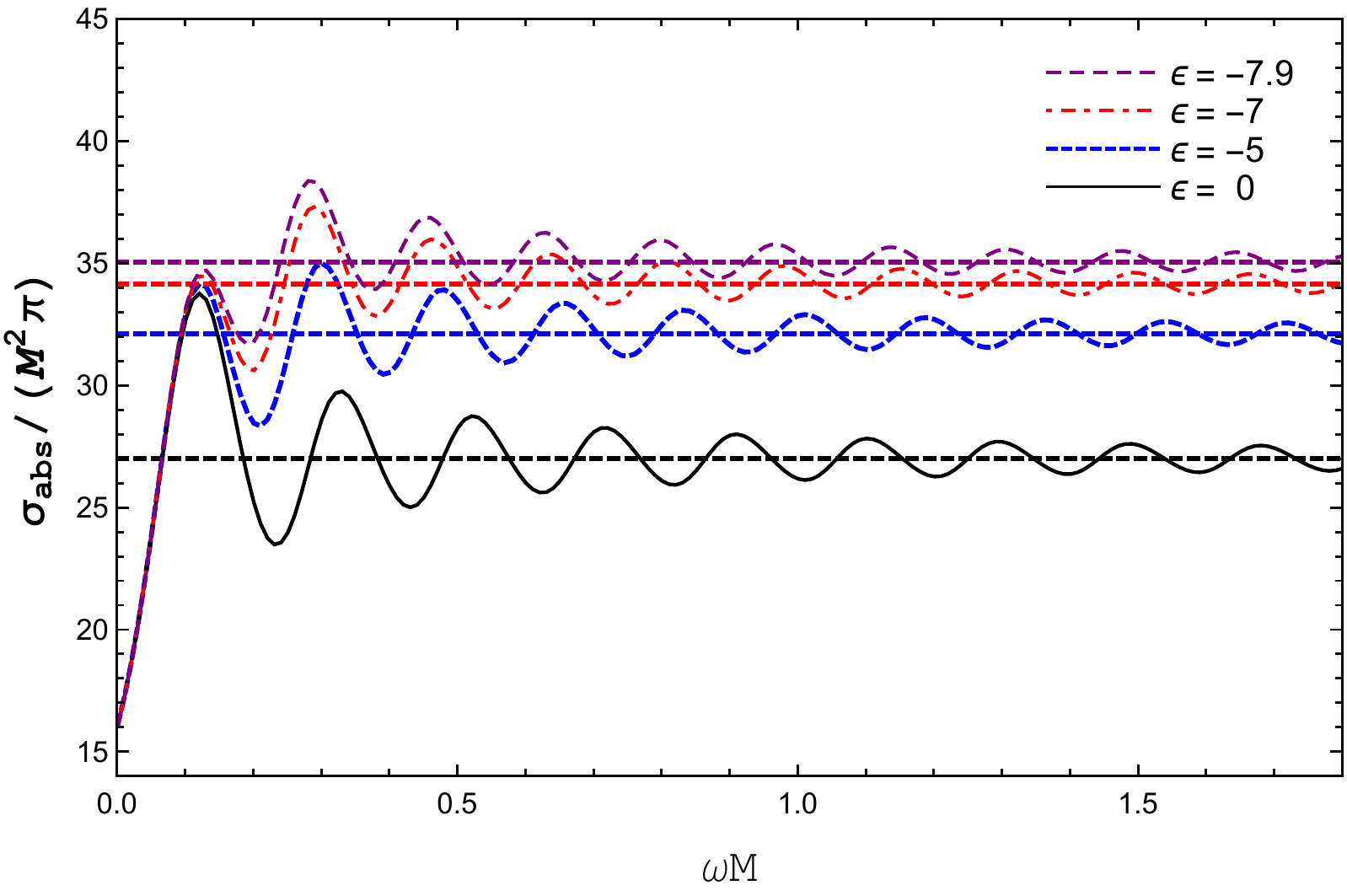}
\caption{Total absorption cross section of JPBHs for some non-positive values of $\epsilon$. Although the first peak ($l=0$) does not change much, the second peak ($l=1$) increases significantly as we decrease the deformation parameter $\epsilon$.}
\label{Schwarzschild1}
\end{figure}

Applying this numerical method we can obtain the transmission coefficients for different values of the deformation parameter, which we plot in Fig.~\ref{transmission}. 
We calculate the partial absorption cross sections for different values of $\epsilon$, by plugging the numerical transmission coefficients into Eq.~\eqref{partialabs}, and plot them in Fig.~\ref{partial}.
In Figs.~\ref{Schwarzschild} and~\ref{Schwarzschild1} we exhibit the total absorption cross sections, obtained by summing over the $l$ modes [cf. Eq.~\eqref{abs}], and compare them with the well-known Schwarzschild case ($\epsilon = 0$). 
We see that as the deformation parameter increases, the total absorption diminishes. 
We also notice, both in Fig.~\ref{Schwarzschild} and in Fig.~\ref{Schwarzschild1} that, for different values of $\epsilon$, the first peak remains almost unaltered, while the other peaks change significantly. 

\begin{figure}[h]
\includegraphics[scale= 0.5]{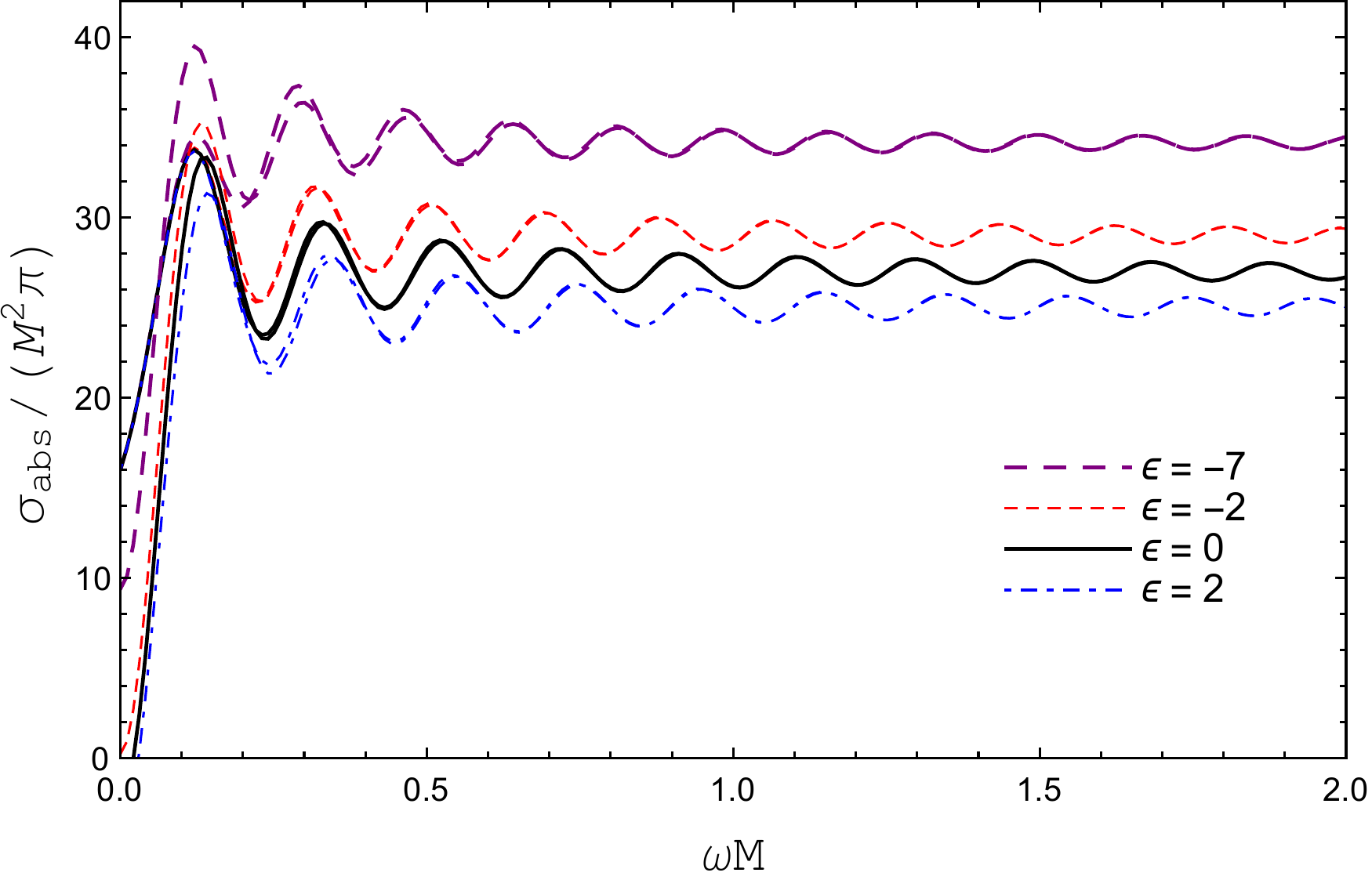}
\caption{Comparison of the sinc approximation with the numerical solution for four values of the deformation parameter ($\epsilon = -7$, $\epsilon = -2$, $\epsilon =0$, and $\epsilon =2$).
}
\label{sincandexact_atual}
\end{figure}

As it can be seen in Fig.~\ref{sincandexact_atual}, 
our numerical results agree very well with the sinc approximation in the high-frequency regime.

For the Schwarzschild case, as pointed out by Unruh in Ref.~\cite{unruh}, the zero-frequency limit is given by the contribution of the lowest mode $l=0$, being equal to the area of the event horizon. We notice that the same happens to the JPBH, independently of the non-vanishing deformation parameter. 
This can also be understood from the behavior of the transmission coefficients of the JPBH (cf.  Fig.~\ref{transmission}), by noticing that, in the low-frequency regime, the transmission coefficient for $l=0$ is dominant. 
Due the dependence of the effective potential~\eqref{eq:effpotential} on the deformation parameter, the radial solution for the lowest mode $l=0$ is the same, independently of the value of $\epsilon$. 
Hence, the transmission coefficient for $l=0$ is the same for any JPBH. Therefore, in the low-frequency regime the total absorption cross section goes to the area of the JPBH, what is consistent with the general result for the scalar absorption of static BHs~\cite{das}.

%%%%%%%%%%%%%%%%%%%%%%%%%%%%%%%%%%%%%%%%%%%%%%%%%
%%%%%%%%%%%%%%%%%%%%%%%%%%%%%%%%%%%%%%%%%%%%%%%%%
\section{Final remarks}\label{sec:remarks}
	%%%%%%%%%%%%%%%%%%%%%%%%%%%%%%%%%%%%%%%%%%%%%%%%%%
	Over the years, several propositions to include new parameters on BH solutions, violating the no-hair theorems, have been presented. In 2011, Johannsen and Psaltis, without requiring the Einstein's equations to be satisfied, proposed a Schwarzschild-like spacetime, regular on and outside the event horizon. 
	The JPBH has an additional parameter, so that two JPBHs with the same mass can deform the spacetime differently.
	
	We have investigated the scalar absorption by JPBHs. We have derived the orbit equation for light rays around JPBHs and solved it to obtain the critical impact parameter and the geometrical absorption cross section, as functions of the deformation parameter. We have shown that if the value of the deformation parameter is increased, both critical impact parameter and geometrical absorption cross section diminishes.
	
	We have used numerical techniques  to compute the partial and total absorption cross sections. We obtained that, as the deformation parameter is increased, the absorption of the scalar field by the JPBH decreases. We also obtained that in the high-frequency regime the total absorption cross section oscillates around the geometrical absorption cross section. 
	In the low-frequency regime, the absorption cross section goes to the area of the black hole, which is independent of the deformation parameter value. 
	
	As a consistency check of our numerical results, we used the sinc approximation to compute the total absorption cross section in the high-frequency regime, obtaining excellent concordance. 
\begin{acknowledgements}
The authors would like to acknowledge 
Conselho Nacional de Desenvolvimento Cient\'ifico e Tecnol\'ogico (CNPq)
and Coordena\c{c}\~ao de Aperfei\c{c}oamento de Pessoal de N\'ivel Superior (CAPES) -- Finance Code 001, from Brazil, for partial financial support. This research has also received funding from the European Union's Horizon 2020 research and innovation programme under the H2020-MSCA-RISE-2017 Grant No. FunFiCO-777740. The authors are grateful to Atsushi Higuchi for profitable discussions.
\end{acknowledgements}

	{}
\end{document}